  \documentclass[final,5p,times,twocolumn]{elsarticle}

\usepackage{epsfig}
\usepackage{rotating}
\usepackage{amssymb}
\usepackage{amsmath}
\usepackage{hyperref}
\usepackage{tabularx}
\usepackage{color}
\journal{Physics Letters B}

\begin{document}

\begin{frontmatter}



\title{Neutrino transition magnetic moments within the non-standard neutrino-nucleus interactions}


\author[label1,label2]{D.K. Papoulias}
\ead{dimpap@cc.uoi.gr}
\author[label1]{T.S. Kosmas}
\ead{hkosmas@uoi.gr}

\address[label1]{Division of Theoretical Physics, University of Ioannina, GR 45110 Ioannina, Greece}
\address[label2]{AHEP Group, Institut de F\'{i}sica Corpuscular --
  C.S.I.C./Universitat de Val\`{e}ncia, Parc Cientific de Paterna.\\
  C/Catedratico Jos\'e Beltr\'an, 2 E-46980 Paterna (Val\`{e}ncia) - Spain}

\begin{abstract}

Tensorial non-standard neutrino interactions are studied through a combined analysis of nuclear structure calculations and a sensitivity $\chi^2$-type of neutrino events expected to be measured at the COHERENT experiment, recently planned to operate at the Spallation Neutron Source (Oak Ridge). Potential sizeable predictions on transition neutrino magnetic moments and other electromagnetic parameters, such as neutrino milli-charges, are also addressed. The non-standard neutrino-nucleus processes, explored from nuclear physics perspectives within the context of quasi-particle random phase approximation, are exploited in order to estimate the expected number of events originating from vector and tensor exotic interactions for the case of reactor neutrinos, studied with TEXONO and GEMMA neutrino detectors. 
\end{abstract}


\end{frontmatter}


\section{Introduction}
\label{Intro}

The investigation of neutrino electromagnetic (EM) properties started long ago \cite{Marciano,Fujikawa}, mainly after the introduction of the minimally extended standard model with right-handed neutrinos \cite{Schechter-Valle}. In this context, at the one loop level the magnetic moment, $\mu_{\nu}$, of a massive neutrino is in general non-zero and its magnitude is proportional to the neutrino mass, $m_\nu$ \cite{Bilenky}. Actually, the theoretical and experimental study of neutrino EM phenomena \cite{Scholberg}, is widely considered as one of the most powerful tools to probe possible interactions involving neutrinos beyond the Standard Model (SM) \cite{PLB,Kosm-talk-Japan}. Furthermore, in an astrophysical environment with extreme conditions (huge magnetic fields, currents, etc.), important non-standard effects may occur due to non-trivial EM properties of neutrinos \cite{Schechter,Giunti}, which may lead to significant alterations of existing scenarios for massive star evolution \cite{Amanik_2005}. 

Exotic neutrino properties arise in neutrino-nucleus processes, occurring due to non-standard neutrino interactions (NSI) of the form \cite{PLB,Kosm-talk-Japan}
\begin{equation}
\nu_{\alpha} (\tilde{\nu}_{\alpha}) + (A,Z) \rightarrow \nu_{\beta} (\tilde{\nu}_{\beta}) + (A,Z) \, ,
\label{neutrin-NSI}
\end{equation}
providing us with model independent constraints of various NSI parameters \cite{Davidson}. In the current literature, even though only vector terms are mainly considered in the relevant Lagrangian \cite{Barranco}, tensorial NSI terms have attracted the interest of studying the aforementioned processes, while robust constraints to the corresponding couplings have been extracted from neutrino-nucleus coherent scattering \cite{Barranco-2012}. In addition, because tensor interaction does not obey the chirality constraint imposed by vector-type couplings, it allows a large class of interactions to be investigated \cite{Healey}. More specifically from a particle physics point of view, tensor NSI terms are possible to be generated via Fierz reordering of the effective low-energy operators appearing in models with scalar leptoquarks \cite{Povarov} as well as in R-parity-violating supersymmetry \cite{Gozdz}.

In general, the non-zero neutrino mass, is experimentally confirmed from neutrino oscillation in propagation data \cite{SK,SNO,KamLAND} and implies that the neutrino is the only particle that exhibits non-standard properties \cite{Shrock-1982}, which are directly connected to the fundamental interactions of particle physics. As a concrete example, neutrino EM properties are useful to distinguish Dirac and Majorana neutrinos and also to probe phenomena of new physics beyond the SM \cite{Bell-2005}. In fact, recent studies, based on model-independent analyses of the contributions to neutrino magnetic moment (NMM), have shown that, if a NMM of the order of $\mu_{\nu} \geq 10^{-15} \mu_{B}$ were experimentally observed, it would confirm the Majorana nature of neutrinos \cite{Bell-2006,Bell-2007}.

The present paper, is an extension of our previous works \cite{PLB, Papoulias-AHEP} in which neutrino-nucleus reactions due to vectorial NSI have been addressed. There, the corresponding couplings have been constrained by exploiting the exceptional sensitivity of the ongoing and planned $\mu^{-} \rightarrow e^{-}$ conversion experiments \cite{Bernstein-Cooper,COMET}. In this Letter, we mainly focus on contributions to the neutrino-nucleus reactions of Eq. (\ref{neutrin-NSI}), due to tensorial terms of the NSI Lagrangian, paying special attention on the nuclear physics aspects of these exotic processes. The cross sections, that arise from the effective four fermion contact interaction Lagrangian, are expressed in terms of the nuclear proton and neutron form factors. Subsequently, the sensitivity on the tensor NSI parameters is obtained from a $\chi^2$ analysis of the expected data from the COHERENT experiment \cite{coherent1,coherent2} recently proposed to operate at the Spallation Neutron Source (SNS) at Oak Ridge \cite{Avignone-Efremenko,Efremenko-2009} by using promising nuclear detectors as $^{20}$Ne, $^{40}$Ar, $^{76}$Ge and $^{132}$Xe. Constraints of this type translate into relevant sensitivities on the upper limits of NMM predicted within the context of the tensor components entering the NSI Lagrangian. The latter can be compared with existing limits derived from $\tilde{\nu}_e-e$ scattering data \cite{Schreckenbach,Kopeikin} coming out of reactor neutrino experiments, such as the TEXONO \cite{TEXONO} and GEMMA \cite{GEMMA} experiments, as well as with other astrophysical observations \cite{Raffelt-1999}. 

On the basis of our nuclear calculations (performed with  quasi-particle RPA) \cite{Kosm-A570,Kosm-2002,Giannaka2} for the dominant coherent process \cite{Chassioti_2009,Balasi}, we evaluate the number of events due to vector and tensor NSI parts of the neutrino-nucleus cross section, and estimate the contribution due to the NMM \cite{Vogel-Engel,Beacom-Vogel,Papageorgiu,PRD}. Our results for the number of events, refer to the $^{76}$Ge isotope which is the current detector medium of the TEXONO and GEMMA experiments. It is worth mentioning that, even though within the SM, gauge invariance and anomaly cancellation constraints require neutrinos to be neutral particles, however non-vanishing electric milli-charge  \cite{Berestetskii} is expected for massive neutrinos which may induce additional neutrino-photon interactions \cite{Studenikin,Gninenko,Chen}.  Furthermore, by taking advantage of the present sensitivity on the transition NMM, we come out with potential stringent constraints (by one order of magnitude more severe than existing limits) on the neutrino milli-charge $q_{\nu}$.

\section{Description of the formalism}
\label{chapt2}
%
\begin{figure}[t]
\begin{center}
\includegraphics[width=0.25\textwidth]{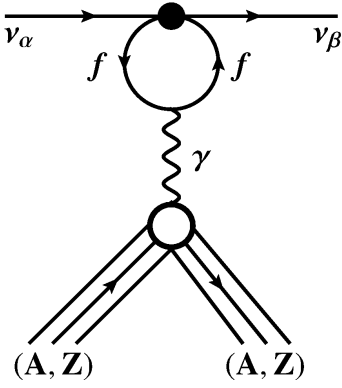}
\end{center}
\caption{Nuclear level effective Feynman diagram for magnetic moment of a neutrino induced by tensorial NSI. The non-standard physics enters in the complicated vertex denoted by the large dot $\bullet$.} 
\label{fig.1}
\end{figure}
%
In general the search for potential existence of phenomena beyond the SM involving NSI at the four fermion approximation, becomes accessible through phenomenological low-energy effective Lagrangians as 
\begin{equation}  \mathcal{L}_{\mathrm{NSI}} = - 2\sqrt{2} G_F \sum_X \sum_{\begin{subarray}{c} f= \, q,\ell\\ \alpha,\beta = \, e,\mu,\tau\end{subarray}} 
\epsilon_{\alpha \beta}^{f X}\left[\bar{\nu}_{\alpha} \Gamma_X \nu_\beta\right]\left[\bar{f} \Gamma_X f\right]\, ,
\label{full_Lagr}
\end{equation}
where $X = \left\{ V,A,S,P,T \right\}$, $\Gamma_X =\left\{ \gamma_\mu, \gamma_\mu \gamma_5, 1, \gamma_5, \sigma_{\mu \nu} \right\} $ and $\sigma_{\mu \nu} = i \left[\gamma_\mu, \gamma_\nu \right]/2$. The magnitude of the NSI couplings $\epsilon_{\alpha \beta}^{f X}$, is taken with respect to the Fermi coupling constant $G_F$ \cite{Scholberg,Barranco},  $\nu_{\alpha}$ denotes three light Majorana neutrinos and $f$ is a quark $q$, or a charged lepton $\ell$. In the present work, we focus on the tensorial $\nu$-nucleus NSI described by the Lagrangian \cite{Barranco-2012}
\begin{equation} \mathcal{L}_{\mathrm{NSI}}^{T} = - 2\sqrt{2} G_F \sum_{\begin{subarray}{c} f= \, u,d\\ \alpha,\beta = \, e,\mu,\tau\end{subarray}} 
\epsilon_{\alpha \beta}^{f T}\left[\bar{\nu}_{\alpha} \sigma^{\mu \nu} \nu_\beta\right]\left[\bar{f} \sigma_{\mu \nu} f\right]\, .
\label{tens_Lagr}
\end{equation}
The extraction of the latter Lagrangian is illustrated in Fig. \ref{fig.1}, where the nuclear-level Feynman loop-diagram represents the photon exchange between a fermion and a quark generating a neutrino magnetic moment. The non-standard physics enters through the complicated leptonic vertex (see also Ref. \cite{PLB}). 
\subsection{Non-standard neutrino-nucleus reaction cross sections}
For neutral current processes, the vector NSI part of the effective Lagrangian (\ref{full_Lagr}), is parametrized in terms of non-universal (NU)  $\epsilon_{\alpha \alpha}^{f V}$ and flavour-changing (FC) vector couplings $\epsilon_{\alpha \beta}^{f V}$ ($\alpha\neq\beta$) \cite{Davidson}.  For coherent scattering, a nucleus of mass $M$ recoils (no intrinsic excitation occurs) with energy which, in the approximation $T_{N} \ll E_{\nu}$ 
(low-energy limit), is maximized as, $T_N^{\text{max}}=2 E_\nu^2/(M+ 2 E_\nu)$. Then, to a good approximation, the square of the three momentum transfer is equal to $q^2 = 2 M T_N$, and the coherent vector NSI differential cross section with respect to $T_{N}$ is written 
as \cite{PLB}
\begin{equation}
\frac{d\sigma_{\mathrm{NSI},\nu_{\alpha}}^{V}}{dT_N} = \frac{G_F^2 \,M}{\pi} \left(1- 
\frac{M\,T_N}{2 E_\nu^2}\right)\left\vert\langle gs\vert\vert 
G_{V,\nu_{\alpha}}^{\mathrm{NSI}} (q) \vert\vert gs \rangle \right \vert ^{2}\, ,
\label{NSI_dT}
\end{equation}
($\alpha = e,\mu,\tau$, denotes the flavour of incident neutrinos) where for even-even nuclei the nuclear ground state reads $\vert gs \rangle=\vert J^\pi \rangle = \vert 0^+ \rangle$. The corresponding nuclear matrix element can be found in Ref. \cite{Papoulias-AHEP}. 

For NSI scattering, the differential cross section with respect to the recoil energy $T_N$ due to tensor interactions (at nuclear level) reads 
\begin{equation}
\frac{d\sigma_{\mathrm{NSI},\nu_{\alpha}}^{T}}{dT_N} = \frac{4 G_F^2 \,M}{\pi} \left[ \left(1- 
\frac{T_N}{2 E_\nu}\right)^2-\frac{M T_N}{4 E_{\nu}^{2}} \right] \left\vert\langle gs\vert\vert 
G_{T,\nu_{\alpha}}^{\mathrm{NSI}} (q) \vert\vert gs \rangle \right \vert ^{2}\, .
\label{tens_dT}
\end{equation}
The corresponding tensorial NSI matrix element arising from the Lagrangian (\ref{tens_Lagr}) takes the form
\begin{equation}
\begin{aligned}
& \left\vert {\cal M}^{\mathrm{NSI}}_{T,\nu_{\alpha}} \right \vert ^{2} \equiv
\left\vert \langle gs \vert \vert G_{T,\nu_{\alpha}}^{\mathrm{NSI}}(q)  \vert \vert gs \rangle \right \vert ^{2}  = \\  & 
 \left[ \left( 2 \epsilon^{uT}_{\alpha \beta} + \epsilon^{dT}_{\alpha \beta} \right) Z F_Z (q^2) +  
\left( \epsilon^{uT}_{\alpha \beta} + 2\epsilon^{dT}_{\alpha \beta} \right) N F_N (q^2) \right]^2 \, ,
\end{aligned}
\label{GT}
\end{equation}
(there is no interference between the tensorial NSI and the SM amplitude \cite{Barranco-2012}) where $F_{Z(N)}(q^2)$ denote the nuclear (electromagnetic) form factors for protons and neutrons. 

\subsection{Neutrino transition magnetic moments}
In flavour space $\alpha,\beta=e,\mu,\tau$, neutrino magnetic moments  $\mu_{\alpha \beta}$ are generated by the tensorial part of the Hermitian magnetic form factor $ f^{M}_{\alpha \beta}(0) = \mu_{\alpha \beta} $ in the effective neutrino EM current \cite{Giunti}
\begin{equation}
-f^{M}_{\alpha \beta}(q^2)\bar{\nu}_{\beta}i \sigma_{\mu \nu} \nu_{\alpha} \, ,
\end{equation}
(for the relation of the NMM between the flavour basis $\mu_{\alpha\beta}$ and the mass basis $\mu_{ij}$ with $i,j=1,2,3$,  see Refs. \cite{Vogel-Engel,Beacom-Vogel}).  It is worth mentioning that, within the minimally extended SM, in order to include neutrino masses, diagonal NMMs $\mu_{\alpha \alpha}$ are possible only for Dirac neutrinos.  However, transition NMMs $\mu_{\alpha\beta}$ can be obtained for both Dirac and Majorana neutrinos \cite{Giunti}.

As it is known, the SM predicts extremely small values for the NMMs (of the order of $\mu_{\nu} \leq 10^{-19}\mu_{B} \left(m_{\nu}/1 \mathrm{eV} \right)$ \cite{Marciano}, where $\mu_{B}$ is the Bohr magneton). Presently, the best upper limit on $\mu_{\nu}$ has been set from astrophysical observations as \cite{Raffelt-1999}
\begin{equation}
\mu_{\nu}\leq 3 \times 10^{-12} \mu_{B} \, .
\label{exp-limit-magn-mom}
\end{equation} 
Other constraints are available through reactor $\tilde{\nu}_{e}-e$ scattering data of the TEXONO experiment \cite{TEXONO}
\begin{equation}
\mu_{\tilde{\nu}_{e}}< 7.4 \times 10^{-11} \mu_{B} \qquad \left(90 \% \,\, \mathrm{C.L.} \right)\, ,
\end{equation}
and of the GEMMA experiment \cite{GEMMA}
\begin{equation}
\mu_{\tilde{\nu}_{e}}< 2.9 \times 10^{-11} \mu_{B} \qquad \left(90 \% \,\, \mathrm{C.L.} \right)\, .
\end{equation}

In our convention the leading order contribution to the NMM for neutrino-quark ($\nu_\alpha-q$) NSI is expressed as
\begin{equation}
\mu_{\alpha \beta}=\sum_{q} 2 \sqrt{2} G_{F} \epsilon_{\alpha \beta}^{qT} \,\frac{N_{c} Q_{q}}{\pi^2} m_{e} m_{q} \, \mathrm{ln} \left( 2 \sqrt{2} G_{F} m_{q}^{2} \right)\, \mu_{B}\, ,
\label{magn-mom}
\end{equation}
where $m_q$ and $Q_q$ are the quark mass and charge respectively, while $N_c$ is the number of quark colours (see also Ref. \cite{Healey}).
Analogously, the NMM for neutrino-lepton ($\nu_\alpha-\ell$) NSI takes the form
\begin{equation}
\mu_{\alpha \beta}= - \sum_{\ell} 2 \sqrt{2} G_{F} \epsilon_{\alpha \beta}^{\ell T} \,\frac{m_{e} m_{\ell}}{\pi^2} \, \mathrm{ln} \left( 2 \sqrt{2} G_{F} m_{\ell}^{2} \right)\, \mu_{B}\, ,
\label{magn-mom2}
\end{equation}
with $m_{\ell}$ being the mass of the charged leptons.

In Ref. \cite{Vogel-Engel}, it has been suggested that the presence of a NMM yields an additional contribution to the weak interaction cross section. Thus, the differential EM cross section $d\sigma/dT_N$ due to a tensor NSI (transition) magnetic moment  is written as
\begin{equation}
\frac{d \sigma_\mathrm{magn}}{dT_N}=\frac{\pi a^2 \mu_{\alpha \beta}^{2}\,Z^{2}}{m_{e}^{2}}\left(\frac{1-T_{N}/E_{\nu}}{T_{N}}+\frac{T_{N}}{4\,E_{\nu}^2}\right) F_{Z}^{2}(q^{2})\,,
\end{equation}
which contains the proton nuclear form factor (see also Ref. \cite{Papageorgiu}). From the Lagrangian (\ref{full_Lagr}) the total cross section reads
\begin{equation}
\label{crossec-full}
\frac{d \sigma_{\mathrm{tot}}}{dT_N} = \frac{d \sigma_{\mathrm{SM}}}{dT_N} + \frac{d \sigma_{\mathrm{NSI}}^{V}}{dT_N} + \frac{d \sigma_{\mathrm{NSI}}^{T}}{dT_N} + \frac{d \sigma_{\mathrm{magn}}}{dT_N} \, ,
\end{equation}
(the flavour indices have been suppressed).
%
\section{Results and discussion}
\subsection{Nuclear structure calculations}

At first, the nuclear structure details that reflect the dependence of the coherent differential cross section on the recoil energy $T_{N}$ through Eq. (\ref{crossec-full}), are studied. This involves realistic calculations of $d\sigma_{\nu_{\alpha}}/dT_N$, for both vector and tensor operators for a set of currently interesting nuclear detectors.  For each nuclear system, the required pairing residual interaction was obtained from a Bonn C-D two-body potential (strong two-nucleon forces) which was slightly renormalized with two pairing parameters $g^{p\,(n)}_{\mathrm{pair}}$  for proton (neutron) pairs \cite{Papoulias-AHEP}. The nuclear form factors for protons and neutrons are obtained as in Ref. \cite{Kosm-A570}, by solving iteratively the BCS equations  \cite{Giannaka2,Chassioti_2009,Balasi}.
\subsection{Tensorial NSI couplings from SNS experiments}
The COHERENT experiment \cite{coherent1,coherent2} proposed to operate at the SNS (Oak Ridge) has excellent capabilities not only to measure, for the first time, coherent neutral-current neutrino-nucleus events, but also to search for new physics beyond the SM \cite{Scholberg}. In general, any deviation from the SM predictions is interesting, therefore in the present study we explore the role of the sensitivity of the above experiment in putting stringent bounds on the tensor NSI, by taking advantage of our realistic nuclear structure calculations. We determine potential limits for the exotic parameters $\epsilon_{\alpha \beta}^{f T}$ and compare them with available constraints reported in similar studies \cite{Barranco-2012,Healey}.

To this aim, we first evaluate the expected number of events, on various detector materials of the COHERENT experiment, through the integral \cite{Scholberg,Papoulias-AHEP}
\begin{equation}
N= K \int_{E_{\nu_\mathrm{min}}}^{E_{\nu_\mathrm{max}}} \eta^{\mathrm{SNS}}(E_{\nu})\,dE_{\nu}\int_{T_{N}^{\mathrm{thres}}}^{T_{N_{\mathrm{max}}}} \frac{d \sigma}{dT_{N}}(E_{\nu},T_{N})\, d T_{N}\, ,
\end{equation}
where $K= N_{targ} \Phi^{\mathrm{SNS}} t_{tot}$, with  $N_{targ}$ being the number of atoms of the studied target nucleus, and $t_{tot}$ the total time of exposure. The relevant neutrino energy distribution $\eta^{\mathrm{SNS}}(E_{\nu})$ and the neutrino fluxes $\Phi^{\mathrm{SNS}}$ (strongly depended on the detector distances from the SNS source), are taken from Refs. \cite{Avignone-Efremenko,Efremenko-2009}.

\begin{figure}[t]
\begin{center}
\includegraphics[width=0.45\textwidth]{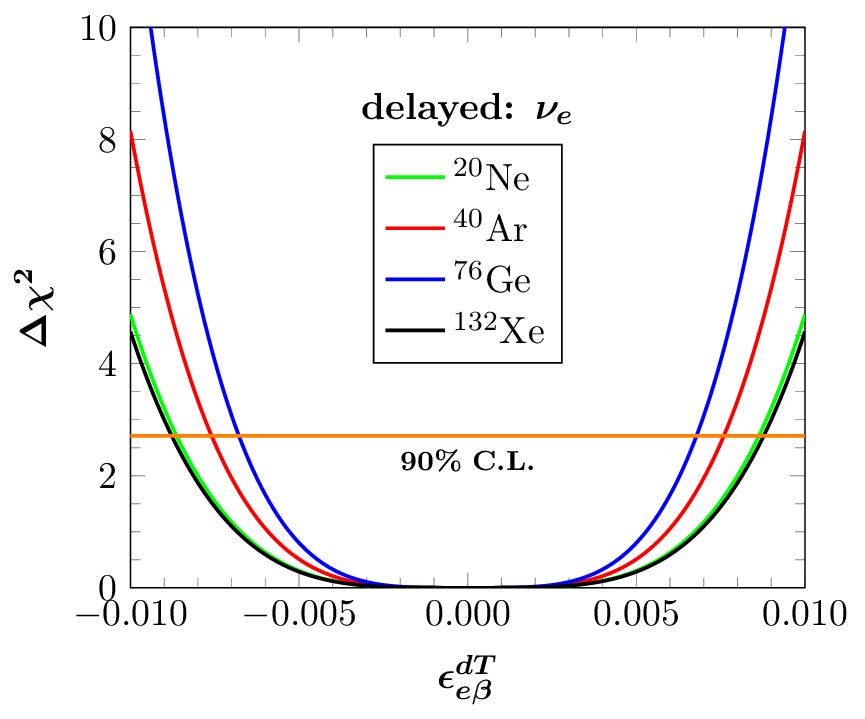}
\end{center}
\caption{$\Delta \chi^2$ profiles as function of the $\epsilon_{e \beta}^{dT}$ NSI parameters, for potential nuclear detectors of the COHERENT experiment (for statistical errors only). }
\label{fig.2}
\end{figure}

To estimate the sensitivity on the tensorial parameters we adopt the futuristic statistical method for the $\chi^2$ defined as \cite{Barranco} 
\begin{equation}
\chi^2 = \left( \frac{N_{events}^{\mathrm{SM}} - N_{events}^{\mathrm{NSI}}}{\delta N_{events}} \right)^2 \, .
\label{chi}
\end{equation}
Since the experiment is not running yet, the calculations are performed without binning the sample relying on statistical errors only (systematic errors are discussed in Ref. \cite{Scholberg}). Calculations which take into consideration possible background errors are addressed in Ref. \cite{PRD}. In Eq. (\ref{chi}) $N_{events}^{\mathrm{SM}}$ ($N_{events}^{\mathrm{NSI}}$) denotes the exact number of SM (tensorial NSI) events expected to be recorded by a COHERENT detector and the  parameters $\epsilon_{\alpha \beta}^{qT}$ are varied so as to fit the hypothetical data.  In our calculations we consider the promising target nuclei, $^{20}$Ne, $^{40}$Ar, $^{76}$Ge, ($^{132}$Xe) at 20 m (40 m) from the SNS source, assuming an energy threshold of 1 keV and a detector mass of one ton. The considered time window of data taking is fixed to one year assuming perfect detection efficiency. For the sake of convenience, from the SNS delayed-beam we take into account only the $\nu_{e}$ component. This allows us also to compare our predictions with those of Ref. \cite{Barranco-2012}. For the various target nuclei, the present results are illustrated in Fig. \ref{fig.2}, from where we conclude that higher  prospects are expected for $^{76}$Ge. In principle, more severe constraints are expected for heavier target nuclei, however, the detector distance from the Spallation target plays crucial role, and thus, a light $^{20}$Ne detector located at 20 m performs better than a heavy $^{132}$Xe detector at 40 m. The corresponding sensitivity at $90\%$ C.L. on the NSI couplings, coming out of the $\nu_{e}$ and the $\tilde{\nu}_{\mu}+\nu_{\mu}$ beams, are listed in Table \ref{table1}. Furthermore, focusing on the $\nu$-quark ($q=u,d$) tensor NSI involved in the Lagrangian (\ref{tens_Lagr}), we exploit the constraints of Table \ref{table1} and utilise Eq. (\ref{magn-mom}), in order to extract the sensitivity on the NMM (see Table \ref{table1}). At this point, we consider useful to make a comparison between the results obtained through our nuclear calculations and those obtained by assuming zero momentum transfer (where $F_{N,Z}(0)=1$) i.e when neglecting the nuclear physics details. This leads to the conclusion that, in the majority of the cases the obtained results differ by about 20\%.
%
%
\begin{table}[t]
\centering
\caption{Constraints on the tensor NSI parameters $\epsilon_{\alpha \beta }^{f T}$ at 90\% C.L. for various potential detector materials of the COHERENT experiment. The sensitivity on transition NMM is also shown at 90\% C.L. }
\label{table1}
\begin{tabular}{{l|cccc}}
\hline
parameter & $^{20}$Ne & $^{40}$Ar & $^{76}$Ge & $^{132}$Xe \\
 \hline \hline
$ |\epsilon_{e \beta}^{d T}| \times 10^{-3}$  & 8.6 & 7.6 & 6.8 & 8.8 \\
$ |\epsilon_{e \beta}^{u T}| \times 10^{-3}$  & 8.6 & 8.1 & 7.5 & 9.8 \\
~$\mu_{e \beta}$~~$\times ~10^{-12} \mu_B$    & 3.0 & 2.7 & 2.5 & 3.2 \\
\hline 
$ |\epsilon_{\mu \beta}^{d T}| \times 10^{-3}$ & 7.1 & 6.3 & 5.6 & 7.2 \\
$ |\epsilon_{\mu \beta}^{u T}| \times 10^{-3}$ & 7.1 & 6.7 & 6.2 & 8.1 \\
~$\mu_{\mu \beta}$~~$ \times ~10^{-12} \mu_B$  & 2.5 & 2.3 & 2.1 & 2.7 \\
\hline
\end{tabular}
\end{table}
%

\begin{figure}[t]
\begin{center}
\includegraphics[width=0.45\textwidth]{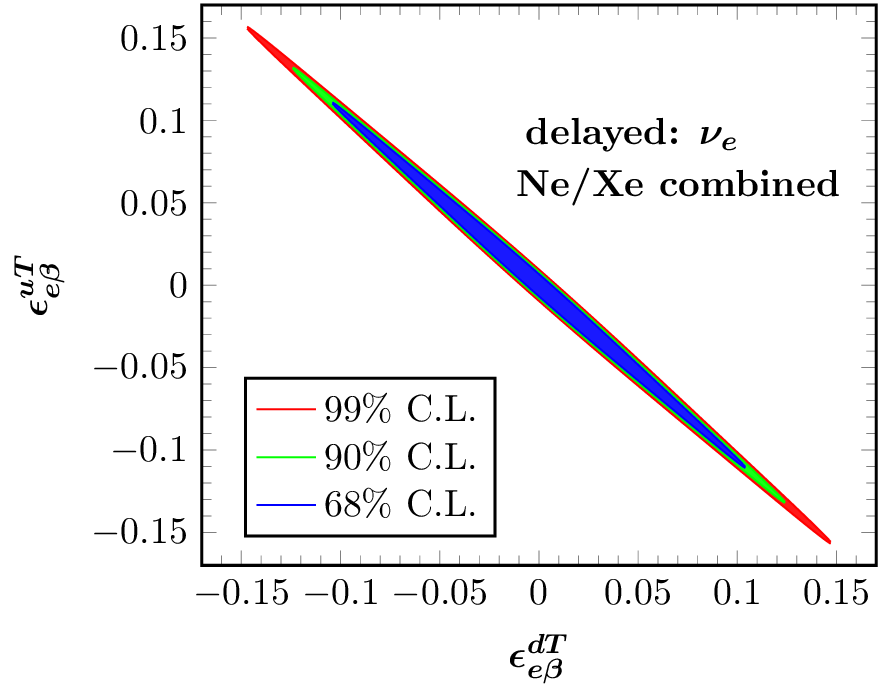}\\
\includegraphics[width=0.45\textwidth]{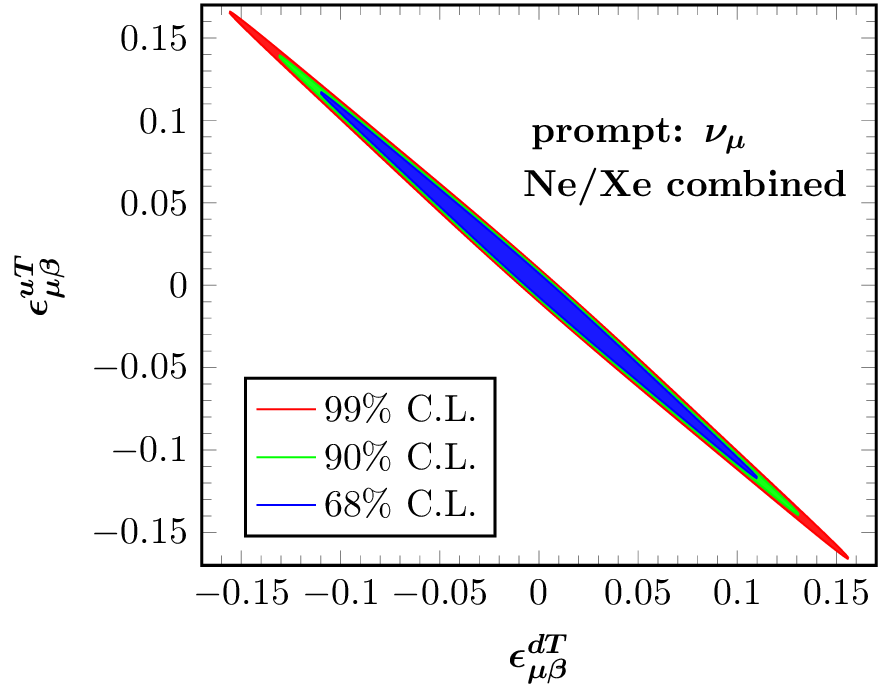}
\end{center}
\caption{Allowed regions in the $\epsilon_{e \beta}^{dT}$-$\epsilon_{e \beta}^{uT}$ (upper panel) and  $\epsilon_{\mu \beta}^{dT}$-$\epsilon_{\mu \beta}^{uT}$ (lower panel) tensor NSI parameter space. Only statistical errors are taken into consideration.} 
\label{fig.3}
\end{figure}

In recent years, it has been shown that, in order to constrain more than one parameters simultaneously, two detectors consisting of target material with maximally different ratio $k=(A+N)/(A+Z)$  are required  \cite{Scholberg,Barranco}. To this purpose, we exploit the advantageous multi-target approach of the COHERENT experiment and in Fig. \ref{fig.3} we illustrate the allowed regions in the $\epsilon_{e \beta}^{dT}$-$\epsilon_{e \beta}^{uT}$ and  $\epsilon_{\mu \beta}^{dT}$-$\epsilon_{\mu \beta}^{uT}$ plane at 68\%,  90\% and 99\% C.L., obtained by varying both tensorial NSI parameters. As expected, the most restricted area corresponds to the delayed beam for which the number of events is larger.
%
%

\subsection{NSI neutrino-nucleus events at the TEXONO experiment}
\begin{figure}[t]
\begin{center}
\includegraphics[width=0.45\textwidth]{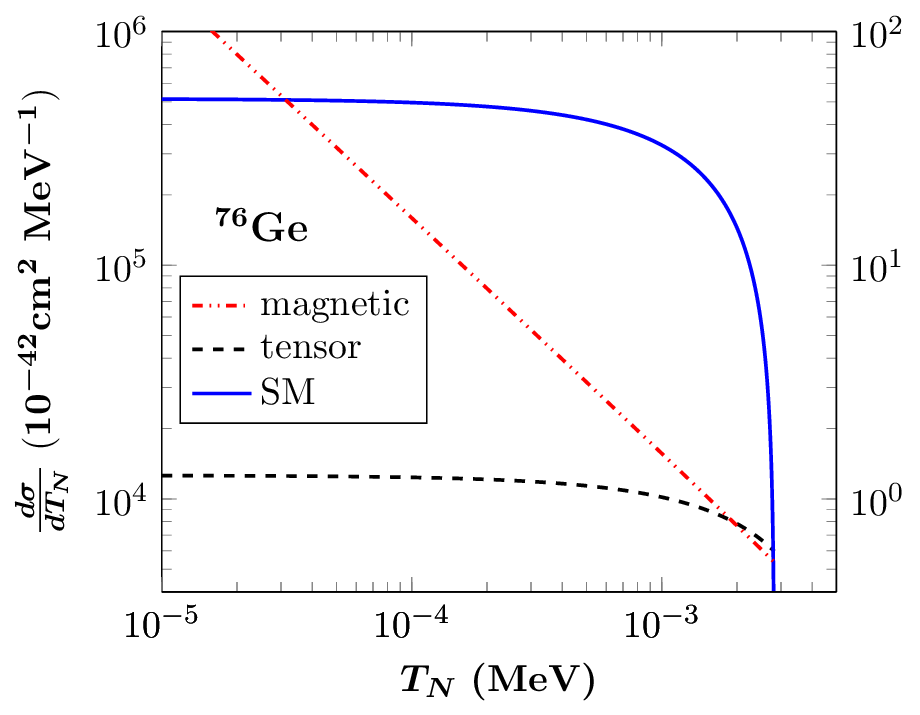}
\end{center}
\caption{Differential cross sections $d\sigma/dT_N$ versus the nuclear recoil energy $T_{N}$, for the SM weak interaction, tensorial NSI and electromagnetic individual parts, assuming ingoing neutrinos with energy 10 MeV. The scale of the EM differential cross section is on the right axis. The utilised parameters for the tensor NSI and the NMM are taken from Table \ref{table1}. }
\label{fig.4}
\end{figure}

One of the most important connections of the present work with ongoing and future reactor neutrino experiments is related to the detection of $\tilde{\nu}_{e}$-nucleus processes.  Towards this aim, for our convenience at first we exploit the available experimental data on the reactor $\tilde{\nu}_{e}$ beams \cite{Schreckenbach,Kopeikin}, in order to fit analytic expressions describing their energy distribution by using numerical optimization techniques. Then, the obtained expressions are applied to predict the number of events expected to be measured in the currently interesting $^{76}$Ge detector material of the TEXONO \cite{TEXONO} and GEMMA \cite{GEMMA} experiments. In Fig. \ref{fig.4} we compare the differential cross sections $d \sigma/ dT_{N}$ for the SM, tensor NSI and electromagnetic components.

From existing measurements of the TEXONO experiment and by employing  Eqs. (\ref{magn-mom}) and (\ref{magn-mom2}), we find the upper bounds on NMM, listed in Table \ref{table2}. Even though some of the derived constraints are less stringent to those given in Table \ref{table1}, it is possible to put limits on more parameters apart from the $\epsilon_{\alpha \beta}^{u(d)T}$.
%
%
%
\begin{table}[t]
\centering
\caption{Upper limits on the NSI parameters $\epsilon_{e \beta }^{f T}$, from
Eq. (\ref{magn-mom}) taking into consideration the results of the TEXONO
experiment. }
\label{table2}
\begin{tabular}{cc|cc|cc}
\hline \hline 
\multicolumn{2}{ c| }{lepton} & \multicolumn{2}{ c| }{quark $Q_q = -1/3$} & \multicolumn{2}{ c }{quark $Q_q = 2/3$} \\
\hline
 $| \epsilon_{e \beta}^{e T} |$  & $3.3 $ & $| \epsilon_{e \beta}^{d T} |$  &
$0.43 $ & $| \epsilon_{e \beta}^{u T} |$  & $0.42$ \\
 $| \epsilon_{e \beta}^{\mu T} |$  & $2.8 \times 10^{-2}$ & $| \epsilon_{e \beta}^{s T} |$  & $2.8 \times 10^{-2}$ & $| \epsilon_{e \beta}^{c T} |$  & $1.7 \times 10^{-3}$ \\
 $| \epsilon_{e \beta}^{\tau T} |$  & $2.7 \times 10^{-3}$ & $| \epsilon_{e
\beta}^{b T} |$  & $1.4 \times 10^{-3}$ & $| \epsilon_{e \beta}^{t T} |$  &
$9.8 \times 10^{-3}$ \\
 \hline \hline
\end{tabular}
\end{table}
%

Since the TEXONO experiment is not running up to now, precise knowledge on the fuel composition is presently not available. For this reason, we focus on the dominant $^{235}$U component of the reactor neutrino distribution covering the energy range $E_{\tilde{\nu}_{e}}<2$ MeV, for which there are only theoretical estimations for the  $\tilde{\nu}_{e}$-spectrum \cite{Kopeikin}. For energies above 2 MeV, we take the existing experimental data from Ref. \cite{Schreckenbach}. In our analysis the normalised spectrum is fitted by the expression 
\begin{equation}
\eta^{\mathrm{react}}_{\tilde{\nu}_{e}}(E_{\nu})=a \left(E_{\nu}\right)^b \, exp\left[c \left(E_{\nu}\right)^d\right]\, ,
\end{equation}
(it resembles the Maxwell-Boltzmann distribution) with the fitted values of parameters: $\alpha=11.36$, $b=1.32$, $c=-3.33$ and $d=0.56$ .
%

\begin{figure}[t]
\begin{center}
\includegraphics[width=0.48\textwidth]{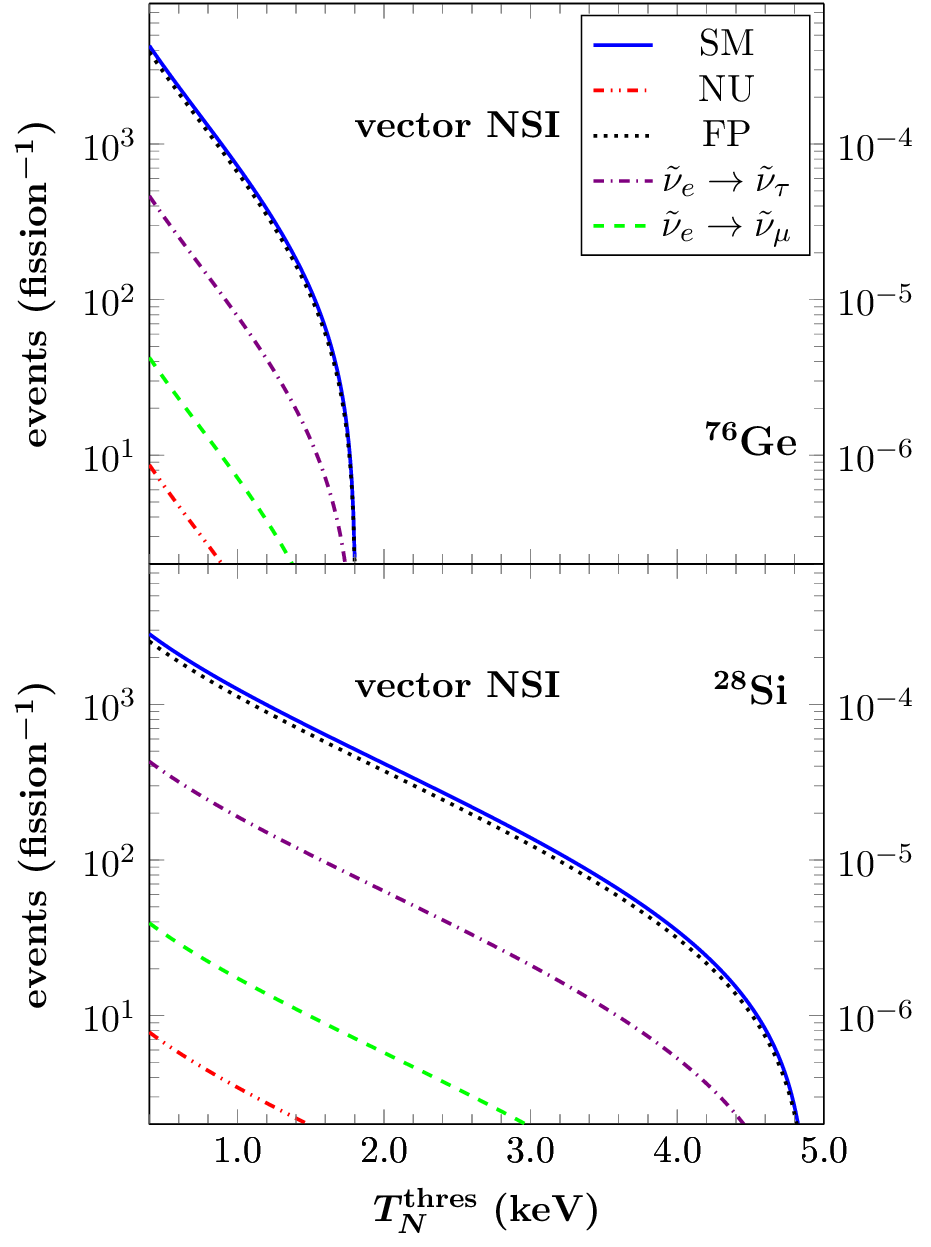}
\end{center}
\caption{Number of events over nuclear recoil threshold due to vectorial NSI, for 1 kg of $^{76}$Ge and 1 kg of $^{28}$Si. The vectorial NSI parameters used, are taken from Refs. \cite{PLB} and \cite{Davidson}. Notice, that the number of counts for the case of the $\nu_e\rightarrow \nu_{\mu}$ reaction channel is plotted with respect to the right axis.}
\label{fig.5}
\end{figure}

\begin{figure}[t]
\begin{center}
\includegraphics[width=0.45\textwidth]{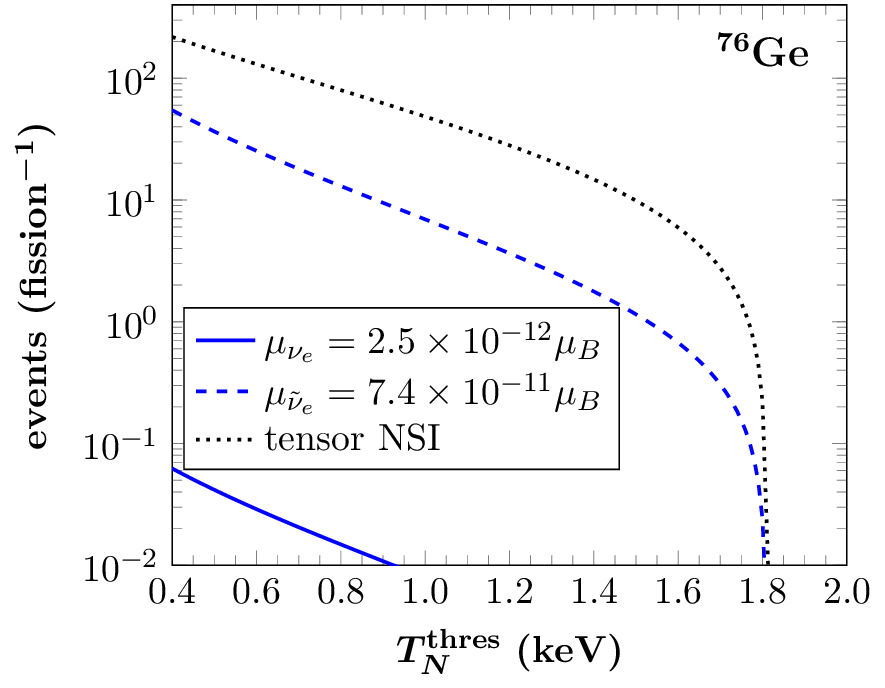}
\end{center}
\caption{Number of events over nuclear recoil threshold due to tensor NSI and NMM for 1 kg of $^{76}$Ge. Constraints for the tensor NSI and NMM parameters are taken from Table \ref{table1}. For comparison, the number of counts due to the NMM using existing limits of the TEXONO experiment, is also illustrated.}
\label{fig.6}
\end{figure}

In Fig. \ref{fig.5}, we present the estimated number of events expected to be measured at the TEXONO experiment, as a function of the nuclear energy-threshold, originating from the various components of the vector NSI.  As detector medium, we consider either 1 kg of $^{76}$Ge or 1 kg of $^{28}$Si, operating with 100\% efficiency for 1 year total exposure, located at 28 m from the reactor core (a typical flux of $\Phi^{\mathrm{react}} = 10^{13}\, \nu \, s^{-1}\ \mathrm{cm^{-2}}$ is assumed). Specifically, for the dominant SM reaction channel, assuming a minimum threshold of $T_{N}^{\mathrm{thres}} = 400 \, \mathrm{eV}$, we find a number of 4280 (2835) scattering events for  $^{76}$Ge ($^{28}$Si) which are in good agreement with previous results \cite{TEXONO}.
%
%
Similarly, in Fig. \ref{fig.6}, we show the total number of counts over threshold due to tensor NSI and NMM, for the same detector composition, by employing the constraints of Table \ref{table1}. Then, for a $^{76}$Ge detector and a $T_{N}^{\mathrm{thres}} = 400 \, \mathrm{eV}$ threshold, our calculations indicate measurable rates, yielding 218 events for processes occurring due to tensor NSI. For interactions due to the presence of a NMM, we obtain $< 1$ events, in comparison to the 55 events expected by incorporating the current TEXONO limit.
%
\subsection{Neutrino milli-charge}
Before closing, we find it interesting to examine the impact of tensorial NSI on other electric properties of the neutrino, that are attributed to the neutrino mass \cite{Vogel-Engel}. Within this framework, milli-charged neutrinos \cite{Gninenko,Chen}, appear with enhanced NMM by acquiring an additional contribution to that which is generated via loop diagrams in theories beyond the SM \cite{Marciano,Bilenky}.  

The differential cross section with respect to the nuclear recoil energy due to an effective neutrino milli-charge, $q_\nu$, is \cite{Berestetskii}
\begin{equation}
\left(\frac{d \sigma}{dT_{e}}\right)_{q_{\nu}} \approx 2 \pi \alpha \frac{1}{m_{e} T_{e}^{2}} q_{\nu}^{2} \, .
\end{equation}
This has to be compared with the magnetic cross section contribution  \cite{Beacom-Vogel}
\begin{equation} 
 \left(\frac{d \sigma}{dT_{e}}\right)_{\mu_{\nu}} \approx \pi \alpha^{2} \frac{1}{m_{e}^{2} T_{e}} \left(\frac{\mu_{\nu}}{\mu_{B}} \right)^{2} \, .
\end{equation} 
 In Ref. \cite{Giunti}, it has been suggested that in order to obtain a limit on the neutrino charge $q_{\nu}$, the ratio $R=\left(\frac{d \sigma}{dT_{e}}\right)_{q_{\nu}} / \left(\frac{d \sigma}{dT_{e}}\right)_{\mu_{\nu}}$ should become smaller than unity, i.e. $R<1$.  Such constraints could be reached irrespectively of whether any deviation from the SM cross section of the $\nu - e^{-}$ process were observed or not \cite{Studenikin}. After writing the upper limit on the neutrino milli-charge in the form
\begin{equation}
\left| q_{\nu} \right| \lesssim 3 \times 10^{-2} \left(\frac{T_{e}}{1 \mathrm{keV}} \right)^{1/2} \left( \frac{\mu_{\nu}}{\mu_{B}}\right) e_{0} \, , 
\end{equation}
and employing the sensitivity on the NMM for the case of $^{76}$Ge (see Table \ref{table1}), for a typical threshold of the order $T_{e}= 400$ eV  we obtain 
\begin{equation}
 \left| q_{\nu} \right| \lesssim 4.7 \times 10^{-14} e_{0}\, .
 \end{equation}
The latter, is by one order of magnitude better than that of previous studies (see Ref. \cite{Giunti}).
%
\section{Summary and Conclusions}

In this work, through the use of realistic nuclear structure calculations, we address various exotic channels of the neutral-current neutrino-nucleus scattering processes. More specifically, we have concentrated on sizeable contributions due to the presence of tensor NSI terms of relevant beyond the SM Lagrangians. Within this framework possible neutrino EM phenomena, that are naturally generated from the tensor operators, such as neutrino transition magnetic moments and neutrino milli-charges, are investigated.

Using our reliable cross sections for SM and NSI $\nu$-processes, we have computed the number of neutrino scattering events expected to be measured at the Spallation Neutron Source experiments. To this purpose, we have chosen as target nuclei the $^{20}$Ne, $^{40}$Ar, $^{76}$Ge and $^{132}$Xe isotopes, that constitute the main detector materials of the planned COHERENT experiment. Through a $\chi^2$-type analysis, we have estimated the sensitivity of the latter experiment on the tensor NSI parameters. We remark, that especially for the case of the $\epsilon_{\mu \beta}^{q T}$ ($q=u,d$) couplings, such bounds are presented here for the first time. Moreover, by exploiting these potential constraints, the resulted sensitivities on the transition neutrino magnetic moments lead to contributions which are of the same order of magnitude with existing limits coming from astrophysical observations. Furthermore, due to their large size, they are accessible by current experimental setups and therefore they may be testable with future experiments searching for coherent neutrino-nucleus scattering. We have also devoted special effort in obtaining precise predictions for the number of neutrino-nucleus events expected to be recorded by the promising TEXONO and GEMMA reactor-$\tilde{\nu}_e$ experiments. 

The present results may contribute usefully towards analysing the detector event-signal, and in conjunction with data expected to be measured in current $\nu$-experiments they may furthermore provide additional information to understand deeper the fundamental electroweak interactions in the neutral- and charged-lepton sector, both for conventional and exotic processes.

\section{Acknowledgements} 
DKP acknowledges financial support by the \textit{Proyecto: " Prometeu per a grups d' investigaci\'o d' Excel$\cdot$l\`{e}ncia de la Conseller\'{i}a d' Educaci\'o, Cultura i Esport, CPI-14-289 "
(GVPROMETEOII2014-084)}. One of us DKP, wishes to thank Dr. Omar Miranda and Dr. Kristopher Healey for stimulating discussions.




\section{References}
\bibliographystyle{elsarticle-num}
\bibliography{<your-bib-database>}

\end{document}